\documentclass[10pt]{iopart}
\usepackage{graphicx}

\begin{document}

\title[Superconducting states of the quasi-2D Holstein model]{Superconducting states of the quasi-2D Holstein model: Effects of vertex and non-local corrections}

\author{J.P.Hague}
\address{Dept. of Physics and Astronomy, University of Leicester, Leicester, LE1 7RH}
\date{22 April 2005}

\begin{abstract}
I investigate superconducting states in a quasi-2D Holstein model
using the dynamical cluster approximation (DCA). The effects of
spatial fluctuations (non-local corrections) are examined and
approximations neglecting and incorporating lowest-order vertex
corrections are computed. The approximation is expected to be valid
for electron-phonon couplings of less than the bandwidth. The phase
diagram and superconducting order parameter are calculated. Effects
which can only be attributed to theories beyond Migdal--Eliashberg
theory are present. In particular, the order parameter shows momentum
dependence on the Fermi-surface with a modulated form and $s$-wave
order is suppressed at half-filling. The results are discussed in
relation to Hohenberg's theorem and the BCS
approximation. {\bf [Published as: J. Phys.: Condens. Matter vol. 17 (2005) 5663-5676]} 
%
%
\end{abstract}

\section{Introduction}
\label{section:introduction}

The discovery of large couplings between electrons and the lattice in
the cuprate superconductors has led to a call for more detailed
theoretical studies of electron-phonon systems in low dimensions
\cite{lanzara2001a,mcqueeny1999a,zhao1997a}. One of the best-known traditional
approaches to the electron-phonon problem is attributed to Migdal and
Eliashberg \cite{migdal1958a,eliashberg1960a}. In a bulk 3D system,
the perturbation theory may be sharply truncated at 1st order and
momentum dependence neglected if the phonon frequency is much less
than the Fermi energy \cite{migdal1958a}. In physical terms, Migdal's
approach requires that there is a very high probability that emitted
phonons are reabsorbed in a last-in-first-out order. The typical
materials of interest at the time were bulk metallic superconductors
where electron-phonon coupling is relatively weak, and the phonon
frequency small compared to the Fermi energy. For this reason, the
application of Migdal--Eliashberg (ME) theory has been very
successful and remains highly regarded.

Strong electron-phonon coupling and large phonon frequencies in low
dimensional systems are outside the limits of validity of the
Migdal--Eliashberg approach. Therefore, the aim of this paper is to
evaluate and discuss the effects of both vertex corrections (VC) and
spatial fluctuations on the theory of coupled electron-phonon systems
in the superconducting state. This follows on from the work by Hague
treating the normal (non-superconducting) state of the Holstein model
using DCA \cite{hague2003a}. Initial attempts to include vertex
corrections were carried out by Engelsberg and Schrieffer
\cite{schrieffera}. Other previous attempts to extend ME theory
include the introduction of vertex corrections into the Eliashberg
equations by Grabowski and Sham \cite{grabowski1984a}, and an
expansion to higher order in the Migdal parameter by Kostur and
Mitrovi\'{c} to investigate the 2D electron-phonon problem
\cite{kostur1993a}. Grimaldi \emph{et al.}  generalised the Eliashberg
equations to include momentum dependence and vertex corrections
\cite{grimaldi1995a}. An anomalous hardening of the phonon mode was
seen by Alexandrov and Schrieffer \cite{schriefferb}. A discussion of
the applicability of these and other approximations to the vertex
function can be found in reference \cite{danylenko2001a}.

The current paper uses DCA to introduce a fully self-consistent
momentum-dependent self-energy. DCA extends DMFT by introducing
short-range fluctuations in a controlled manner
\cite{hettler1998a}. It is particularly good at describing the
electron-phonon problem, due to the limited momentum dependence of the
self-energy, and in this case, the self-consistent DCA can be viewed
as an expansion about the Eliashberg equations (in which momentum
dependence is effectively coarse grained in a manner similar to DMFT)
\cite{eliashberg1960a}. In contrast to the Eliashberg equations, the
full form of the Green's function is considered here, rather than the
renormalised weak coupling Green's function (which has the form
$\underline{G}^{-1}(\epsilon_k,i\omega_n)=Z i\omega_n
\underline{\sigma}_0 - (\epsilon_k+\chi) \underline{\sigma}_3 - \Delta
\underline{\sigma}_1$).

Two approximations for the electron and phonon self energies are
applied in this paper. The first neglects vertex corrections, but
incorporates non-local fluctuations. The second incorporates lowest
order vertex and non-local corrections. The vertex corrections allow
the sequence of phonon absorption and emission to be reordered once,
and therefore introduce exchange effects. The DCA result is compared
to the corresponding DMFT result and in this way low-dimensional
effects are isolated. It should be noted that in the extreme strong
coupling limit, the Holstein model forms a bipolaronic ground state,
and perturbative methods in the electron and phonon Green's functions
break down \cite{alexandrovandmott}. In the dilute limit, the Holstein
model forms a polaronic liquid. There are significant differences
between the weak and strong coupling limits of the polaron problem. In
the strong-coupling limit, the Lang--Firsov approximation may be
applied, and physical properties have very different behaviour. For
example the effective mass is reduced by an exponential factor of
coupling. Exact numerical results show that the crossover between weak
and strong coupling regimes occurs rapidly at $\lambda\sim 1$
\cite{kornilovitch}. For this reason, the present approximation should
be considered valid for $|U|<W$.

The paper is organised as follows. In section \ref{section:dca}, the
DCA is introduced. In section \ref{section:holstein}, the Holstein
model of electron-phonon interactions is described, and the
perturbation theory and the full algorithm used in this work are
detailed. In section \ref{section:results}, the results are
presented. The momentum dependence of the superconducting order
parameter is examined through the density of superconducting
pairs. The phase diagram is then computed and comparison is made with
analytical results. A summary of the major findings of this research
is provided in section \ref{section:summary}.

\section{The dynamical cluster approximation}
\label{section:dca}

\begin{figure}[t]
\begin{indented}\item[]
\includegraphics[width=60mm]{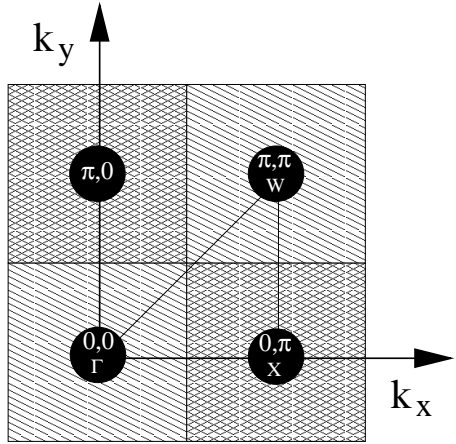}
\end{indented}
\caption{A schematic representation of the reciprocal-space coarse
graining scheme for a 4 site DCA. Within the shaded areas, the
self-energy is assumed to be constant. There is a many to one mapping
from the crosshatched areas to the points at the centre of those
areas. The coarse-graining procedure corresponds to the mapping to a
periodic cluster in real space, with spatial extent
$N_{c}^{1/D}$. Also shown are the high symmetry points $\Gamma$, $W$
and $X$, and lines connecting the high symmetry points. An infinite
number of $\mathbf{k}$ states are involved in the coarse-graining
step, so the approximation is in the thermodynamic limit. DMFT
corresponds to $N_{C}=1$.}
\label{fig:coarsegrain}
\end{figure}

The dynamical cluster approximation \cite{hettler1998a,hettler2000a}
is an extension to the dynamical mean-field theory. DMFT has been
applied as an approximation to models of 3D materials
\cite{georges1996a,tahvildar1997a,miller1998a}. However, application
of DMFT to one- and two-dimensional models gives an incomplete
description of the physics. An example of significant differences
between two- and three-dimensional physics comes from quantum
spin-systems. In 3D, the Heisenberg model orders at a transition
temperature, $T_N$. Significant non-local fluctuations in two
dimensions reduce the N\'{e}el temperature to zero (Mermin--Wagner
theorem), and the mean-field approach fails completely.

Conceptually, DCA is similar to DMFT. The Brillouin zone is divided up
into $N_C$ subzones consistent with the lattice symmetry (see figure
\ref{fig:coarsegrain}). Within each of these zones, the self-energy is
assumed to be momentum independent. For a system in the normal state,
the Green's function is determined as,
\begin{equation}
\label{eqn:dcagreensfunction}
G(\mathbf{K}_{i},z)=\int _{-\infty }^{\infty }\frac{{\mathcal{D}}_{i}(\epsilon )\, d\epsilon }{z+\mu-\epsilon -\Sigma (\mathbf{K}_{i},z)}
\end{equation}
where \( {\mathcal{D}}_{i}(\epsilon ) \) is the non-interacting
Fermion density of states for subzone $i$, and the vectors
$\mathbf{K}_i$ represent the average $\mathbf{k}$ for each subzone
(plotted as the large dots in figure \ref{fig:coarsegrain}). The
theory deals with the thermodynamic limit, and introduces non-local
fluctuations with a characteristic length scale of $N_C^{1/D}$. For
$N_C=1$, DCA is equivalent to DMFT.

Since superconducting states are to be considered, DCA is extended
within the Nambu formalism \cite{maier2004} in a similar manner to DMFT
\cite{georges1996a}. Green's functions and self-energies are
described by $2\times 2$ matrices, with off-diagonal anomalous terms
relating to the superconducting states. Note that in the following
equations 4-vectors are used, i.e. $\mathbf{K}\equiv
(i\omega_n,\mathbf{K})$. The Green's function and self-energy matrices
have the components,
\begin{equation}
\underline{G}(\mathbf{K})=\left( \begin{array}{cc} G(\mathbf{K}) & F(\mathbf{K}) \\ F^{*}(\mathbf{K}) & -G(-\mathbf{K}) \end{array} \right)
\label{eqn:gfmatrix}
\end{equation}
\begin{equation}
\underline{\Sigma}(\mathbf{K})=\left( \begin{array}{cc} \Sigma(\mathbf{K}) & \phi(\mathbf{K}) \\ \phi^{*}(\mathbf{K}) & -\Sigma(-\mathbf{K}) \end{array} \right)
\label{eqn:sematrix}
\end{equation}

The coarse graining step is generalised to the superconducting state
as,
\begin{equation}
G(\mathbf{K},i\omega_n)=\int_{-\infty}^{\infty}d\epsilon\frac{{\mathcal{D}}_i(\epsilon)(\zeta(\mathbf{K}_i,i\omega_n)-\epsilon)}{|\zeta(\mathbf{K}_i,i\omega_n)-\epsilon|^2+\phi(\mathbf{K}_i,i\omega_n)^{2}}
\label{eqn:grnsfnsc}
\end{equation}
\begin{equation}
F(\mathbf{K},i\omega_n)=-\int_{-\infty}^{\infty}d\epsilon\frac{{\mathcal{D}}_i(\epsilon)\phi(\mathbf{K}_i,i\omega_n)}{|\zeta(\mathbf{K}_i,i\omega_n)-\epsilon|^2+\phi(\mathbf{K}_i,i\omega_n)^{2}}
\label{eqn:grnsfnscanom}
\end{equation}
where
$\zeta(\mathbf{K}_i,i\omega_n)=i\omega_n+\mu-\Sigma(\mathbf{K}_i,i\omega_n)$.

The symmetry of the problem was constrained using the $\mathrm{p}m3m$
\emph{planar} point group suitable for a 2D square lattice
\cite{internationaltables}. The partial DOS used in the
self-consistent condition were calculated using the analytic
tetrahedron method to ensure very high accuracy \cite{lambin1984a}.

\section{The Holstein model}
\label{section:holstein}

\begin{figure}[t]
\begin{indented}\item[]
\includegraphics[width=80mm]{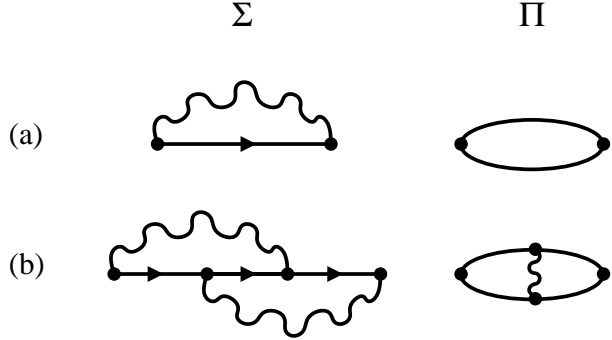}
\end{indented}
\caption{Diagrammatic representation of the approximation used in this
paper. Series (a) represents the vertex-neglected theory which
corresponds to the Migdal--Eliashberg approach. This is valid when
there is a high probability that the last emitted phonon is the first
to be reabsorbed, which is true if the phonon energy $\omega_0$ and
electron-phonon coupling $U$ are small compared to the Fermi
energy. Series (b) represents additional diagrams for the vertex
corrected theory. The inclusion of the lowest order vertex correction
allows the order of absorption and emission of phonons to be swapped
once. For moderate phonon frequency and electron-phonon coupling,
these additions to the theory, in combination with non-local
corrections are expected to improve the theory to sufficient
accuracy. The phonon self energies are labeled with $\Pi$, and
$\Sigma$ denotes the electron self-energies. Lines represent the full
electron Green's function and wavy lines the full phonon Green's
function.}
\label{fig:feynmandiag}
\end{figure}

A simple, yet non-trivial, model of electron-phonon interactions
treats phonons as nuclei vibrating in a time-averaged harmonic
potential (representing the interactions between all nuclei) i.e. only
one frequency $\omega_0$ is considered. The phonons couple to the
local electron density via a momentum-independent coupling
constant $g$. The resulting \emph{Holstein Hamiltonian}
\cite{holstein1959} is written as,
\begin{equation}
H=-\sum_{ij\sigma}t_{<ij>\sigma}c^{\dagger}_{i\sigma}c_{j\sigma}+\sum_{i\sigma} n_{i\sigma} (gr_i-\mu)+\sum_i\left( \frac{M\omega_{0}^2r_i^2}{2}+\frac{p_i^2}{2M}\right)
\end{equation}
The first term in this Hamiltonian represents a tight binding model
with hopping parameter $t$. Its Fourier transform takes the form
$\epsilon_{k}=-2t\sum_{i=1}^{D}\cos(k_{i})$. The second term connects
the local ion displacement, \( r_{i} \) to the local electron
density. Finally the last term can be identified as the bare phonon
Hamiltonian, which is a simple harmonic oscillator. The creation and
annihilation of electrons is represented by $c^{\dagger }_{i}$ and
$c_{i}$ respectively, \( p_{i} \) is the ion momentum and \( M \) the
ion mass. $t=0.25$ in this paper, corresponding to a bandwidth of
$W=2$. A small interplanar hopping of $t_{\perp}=0.01$ is included to
reduce the strength of the logarithmic singularity at $\epsilon=0$ in
${\mathcal{D}}_{\pi,0}(\epsilon)$ and
${\mathcal{D}}_{0,\pi}(\epsilon)$ and stabilise the solution. This is
only expected to modify the results at very low temperature for large
clusters, and gives the problem a quasi-2D character.

It is possible to find an expression for the effective interaction between
electrons by integrating out phonon degrees of freedom \cite{bickers1989}. In
Matsubara space, this interaction has the form,
\begin{equation}
\label{eqn:phononfn}
U(i\omega _{s})=\frac{U\omega_{0}^{2}}{\omega _{s}^{2}+\omega _{0}^{2}}
\end{equation}
Here, $\omega _{s}=2\pi sT$ represent the Matsubara frequencies for
Bosons and $s$ is an integer. A variable $U=-g^2/M\omega_0^2$ is
defined to represent the effective electron-electron coupling in the
remainder of this paper.

When phonon frequency and coupling are small, Migdal's theorem
applies. Migdal's approach allows vertex corrections to be neglected
and becomes exact when $U\rightarrow 0^{-}$, $\omega_0\rightarrow 0^+$
and is 1st order in $U$. In the limit of huge phonon frequency, the
model maps onto an attractive Hubbard model, so the weak coupling
limit of the Holstein model is only obtained by considering all
second-order diagrams in $U$, and ME theory fails. The
vertex-corrected theory described in this paper has the appropriate
weak coupling behaviour for both large and small $\omega_0$.

In this paper, perturbation theory to 2nd order in $U$ is used
\cite{miller1998a} (figure \ref{fig:feynmandiag}). The derivation of
the perturbation theory in Ref. \cite{miller1998a} made use of the
conserving approximations of Bahm and Kadanoff
\cite{bahm,bickers1989}, which Miller \emph{et al.} then simplified
by applying the dynamical mean-field theory (or local
approximation). Here the theory has been extended to include partial
momentum dependence through the application of the DCA. The electron
self-energy has two terms,
$\Sigma_{\mathrm{ME}}(\mathbf{K},i\omega_n)$ neglects vertex
corrections (figure \ref{fig:feynmandiag}(a)), and
$\Sigma_{\mathrm{VC}}(\mathbf{K},i\omega_n)$ corresponds to the vertex
corrected case (figure
\ref{fig:feynmandiag}(b)). $\Pi_{\mathrm{ME}}(\mathbf{K},i\omega_s)$
and $\Pi_{\mathrm{VC}}(\mathbf{K},i\omega_s)$ correspond to the
equivalent phonon self energies. The diagrams translate as follows:

\begin{equation}
\Sigma_{\mathrm{ME}}(\mathbf{K})=UT\sum_{\mathbf{Q}}G(\mathbf{Q})D(\mathbf{K}-\mathbf{Q})
\label{eqn:selfenergyme}
\end{equation}
\begin{equation}
\phi_{\mathrm{ME}}(\mathbf{K})=-UT\sum_{\mathbf{Q}}F(\mathbf{Q})D(\mathbf{K}-\mathbf{Q})
\label{eqn:anomse}
\end{equation}
\begin{equation}
\Pi_{\mathrm{ME}}(\mathbf{K})=-2UT\sum_{\mathbf{Q}}[G(\mathbf{Q})G(\mathbf{K}+\mathbf{Q})-F(\mathbf{Q})F^{*}(\mathbf{K}+\mathbf{Q})]
\end{equation}
\begin{eqnarray}
\Sigma_{\mathrm{VC}}(\mathbf{K})=(UT)^2\sum_{\mathbf{Q}_1,\mathbf{Q}_2}[& G(\mathbf{Q}_1)G(\mathbf{Q}_2)G(\mathbf{K}-\mathbf{Q}_2-\mathbf{Q}_1)\nonumber\\
&-F(\mathbf{Q}_1)G(\mathbf{Q}_2)F^{*}(\mathbf{K}-\mathbf{Q}_2-\mathbf{Q}_1)\nonumber\\
&-F^{*}(\mathbf{Q}_1)G(\mathbf{Q}_2)F(\mathbf{K}-\mathbf{Q}_2-\mathbf{Q}_1)\nonumber\\
&-G^{*}(\mathbf{Q}_1)F(\mathbf{Q}_2)F^{*}(\mathbf{K}-\mathbf{Q}_2-\mathbf{Q}_1)]\nonumber\\
& \times D(\mathbf{K}-\mathbf{Q}_2)D(\mathbf{Q}_1-\mathbf{Q}_2)
\end{eqnarray}
\begin{eqnarray}
\phi_{\mathrm{VC}}(\mathbf{K})=(UT)^2\sum_{\mathbf{Q}_1,\mathbf{Q}_2}[& F^{*}(\mathbf{Q}_1)F(\mathbf{Q}_2)F(\mathbf{K}-\mathbf{Q}_2+\mathbf{Q}_1)\nonumber\\
&-G(\mathbf{Q}_1)F(\mathbf{Q}_2)G(\mathbf{K}-\mathbf{Q}_2+\mathbf{Q}_1)\nonumber\\
&-G^{*}(\mathbf{Q}_1)F(\mathbf{Q}_2)G^{*}(\mathbf{K}-\mathbf{Q}_2+\mathbf{Q}_1)\nonumber\\
&-F(\mathbf{Q}_1)G(\mathbf{Q}_2)G^{*}(\mathbf{K}-\mathbf{Q}_2+\mathbf{Q}_1)]\nonumber\\
& \times D(\mathbf{K}-\mathbf{Q}_2)D(\mathbf{Q}_1-\mathbf{Q}_2)
\end{eqnarray}
\begin{eqnarray}
\Pi_{\mathrm{VC}}(\mathbf{K})=-(UT)^2\sum_{\mathbf{Q}_1,\mathbf{Q}_2}\mathrm{Tr}&\left\{\underline{\sigma}_3\underline{G}(\mathbf{Q}_2+\mathbf{K})\underline{\sigma}_3\underline{G}(\mathbf{Q}_2)\underline{\sigma}_3\underline{G}(\mathbf{Q}_1)\underline{\sigma}_3\underline{G}(\mathbf{K}+\mathbf{Q}_1)\right\}\nonumber\\
&\times D(\mathbf{Q}_2-\mathbf{Q}_1)
\end{eqnarray}
where $\underline{\sigma}_3$ is the third Pauli
matrix. $\underline{\Sigma}=\underline{\Sigma}_{\mathrm{ME}}+\underline{\Sigma}_{\mathrm{VC}}$
and $\Pi=\Pi_{\mathrm{ME}}+\Pi_{\mathrm{VC}}$.

The coarse-grained phonon propagator $D(\mathbf{K},i\omega_s)$ is
calculated from,
\begin{equation}
D(\mathbf{K},i\omega_s)=\frac{\omega_0^2}{\omega_s^2+\omega_0^2-\Pi(\mathbf{K},i\omega_s)}
\label{eqn:phonprop}
\end{equation}
since the bare dispersion of the Holstein model is flat.

The time taken to perform the double integration over momentum and
Matsubara frequencies is the main barrier to performing
vertex-corrected calculations, and this limits the cluster size. Since
the Holstein model with $\omega_0,|U| \ll W$ ($W$ is the bandwidth)
has fluctuations which are almost momentum independent, the DCA has
especially fast convergence in $N_C$ for the parameter regime where
$\omega_0,|U|< W$, and calculations with relatively small cluster size
accurately reflect the physics \cite{hague2003a}. In this respect,
finite size calculations take too long to compute, and the application
of DCA to this problem is essential.

\section{Results}
\label{section:results}

In this section, I discuss results from the self-consistent
scheme. Calculations are carried out along the Matsubara axis, with
sufficient Matsubara points for an accurate calculation. The vertex
corrected self-energies drop off more quickly with Matsubara
frequency, so it is possible to increase efficiency by calculating for
less frequencies. Typically, 256 Matsubara frequencies are used for
the vertex neglected diagrams, and 64 for the vertex corrected
diagrams, which reach asymptotic behaviour at smaller Matsubara
frequencies. The scheme was iterated until the normal and anomalous
self-energies had converged to an accuracy of approximately 1 part in
$10^4$. This corresponds to a very high accuracy for the Green's
function.

Obtaining superconducting solutions involves an additional step, which
is not obvious at the outset. Since the anomalous Green's function is
proportional to the anomalous self energy, initialising the problem
with the non-interacting Green's function leads to a
non-superconducting (normal) state. Also, the non-interacting Green's
function is consistent with an ungapped state and opening a gap in the
electron spectrum can lead to limit cycles during self-consistency,
which are damped in the normal way \cite{georges1996a}.

To induce superconductivity, a constant superconducting field is
applied to the whole system, leading to a non-zero anomalous Green's
function, and automatically opening a gap in the normal-state Green's
function. The procedure of applying a fictitious superconducting field
is analogous to the application of a magnetic field to a spin system
to induce a moment (the order parameter in that case). The
superconducting field is applied by adding a constant term to the
anomalous self-energy in equation \ref{eqn:anomse}. With the field
applied, equations \ref{eqn:grnsfnsc},\ref{eqn:grnsfnscanom} and
\ref{eqn:selfenergyme}-\ref{eqn:phonprop} are solved self-consistently
until convergence is reached. \emph{Once satisfactory convergence is
reached, the fictitious field is completely removed}. Iteration then
continues until the true superconducting state is reached. This
procedure corresponds to initialising the self-consistent cycle with a
superconducting solution; note that similar techniques are used for
obtaining Mott insulating solutions in the Hubbard model using DMFT
\cite{georges1996a}.

By following this procedure, a superconducting state may be found
below the transition temperature, $T_C$. Green's functions and self-energies
computed in the superconducting state can then be used to initialise
the self-consistent equations for similar couplings, fillings,
temperatures (with an appropriate rescaling of the Matsubara
frequencies) and phonon frequencies. Above the transition temperature,
the magnitude of the anomalous Green's function tends to zero during
self-consistency as expected.

\begin{figure}
\begin{indented}\item[]
\includegraphics[width=35mm,angle=270]{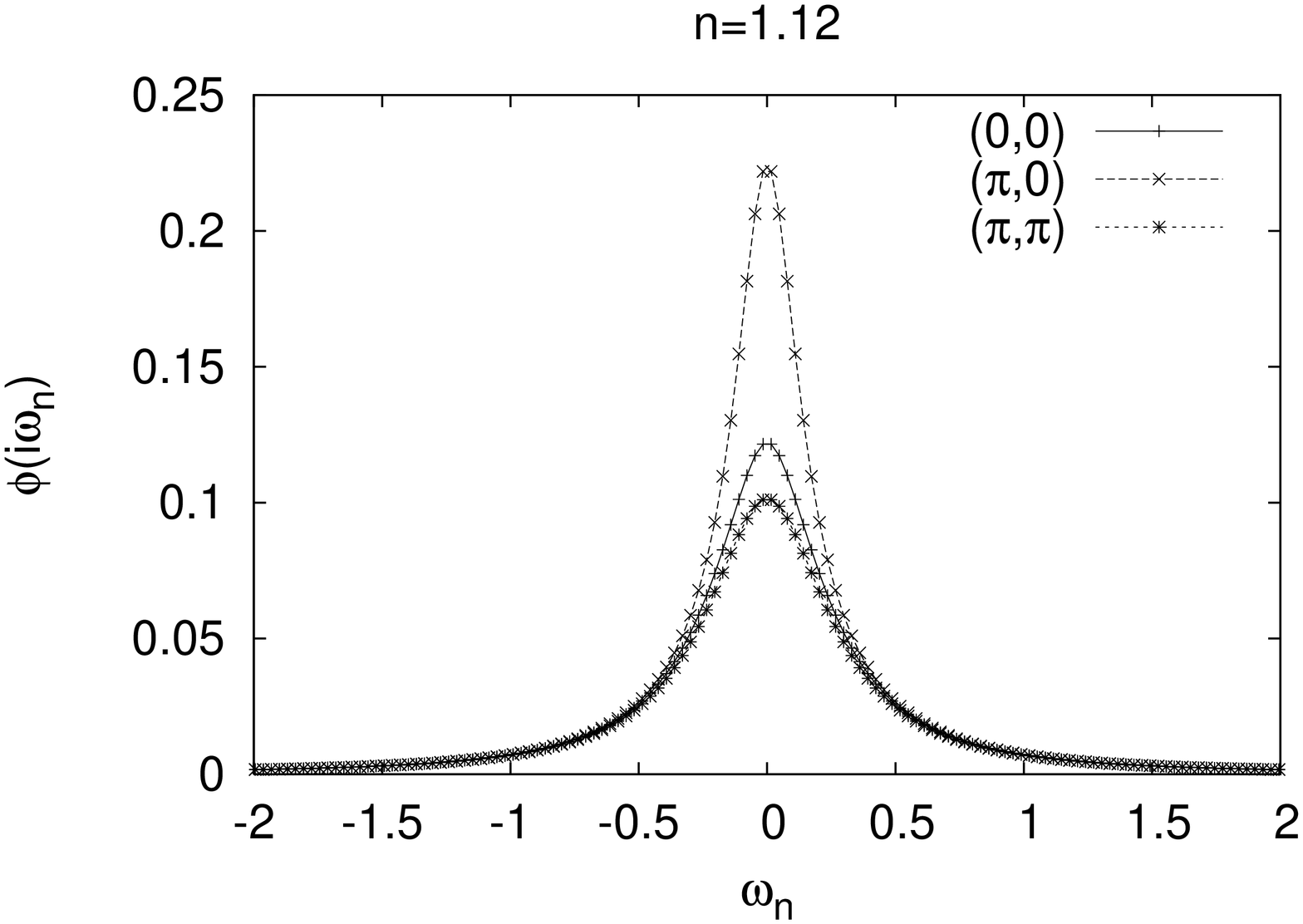}
\includegraphics[width=35mm,angle=270]{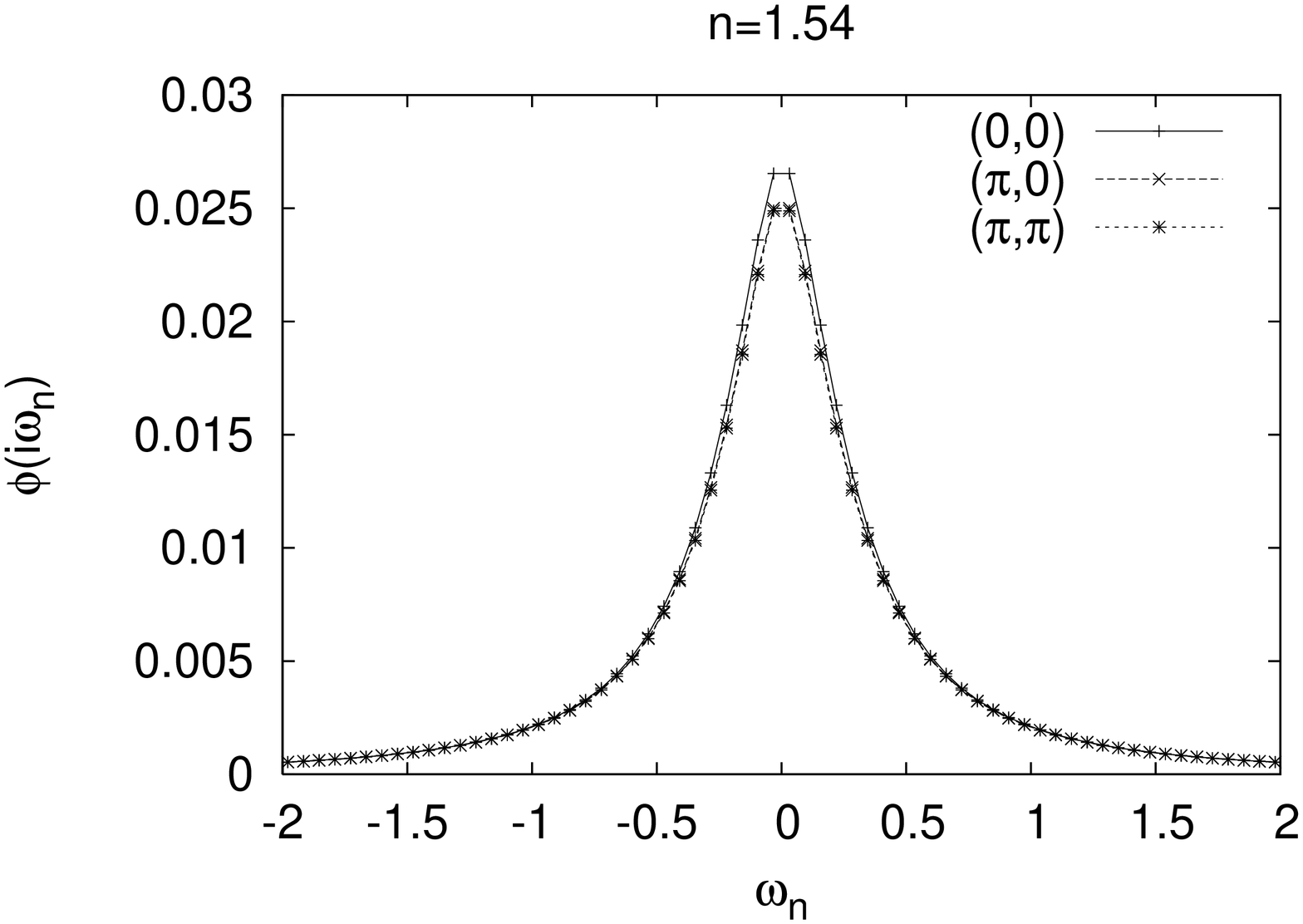}
\includegraphics[width=35mm,angle=270]{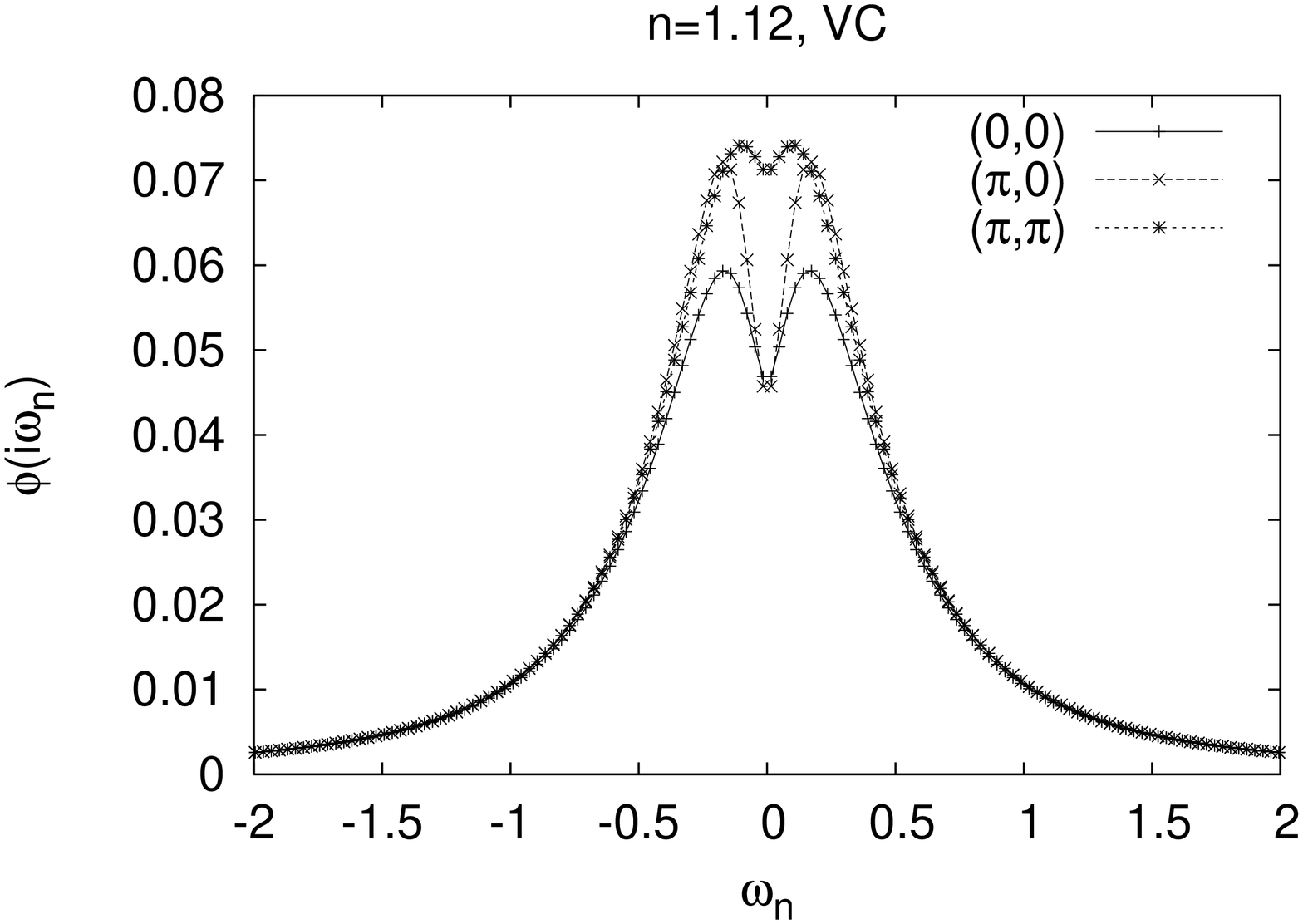}
\includegraphics[width=35mm,angle=270]{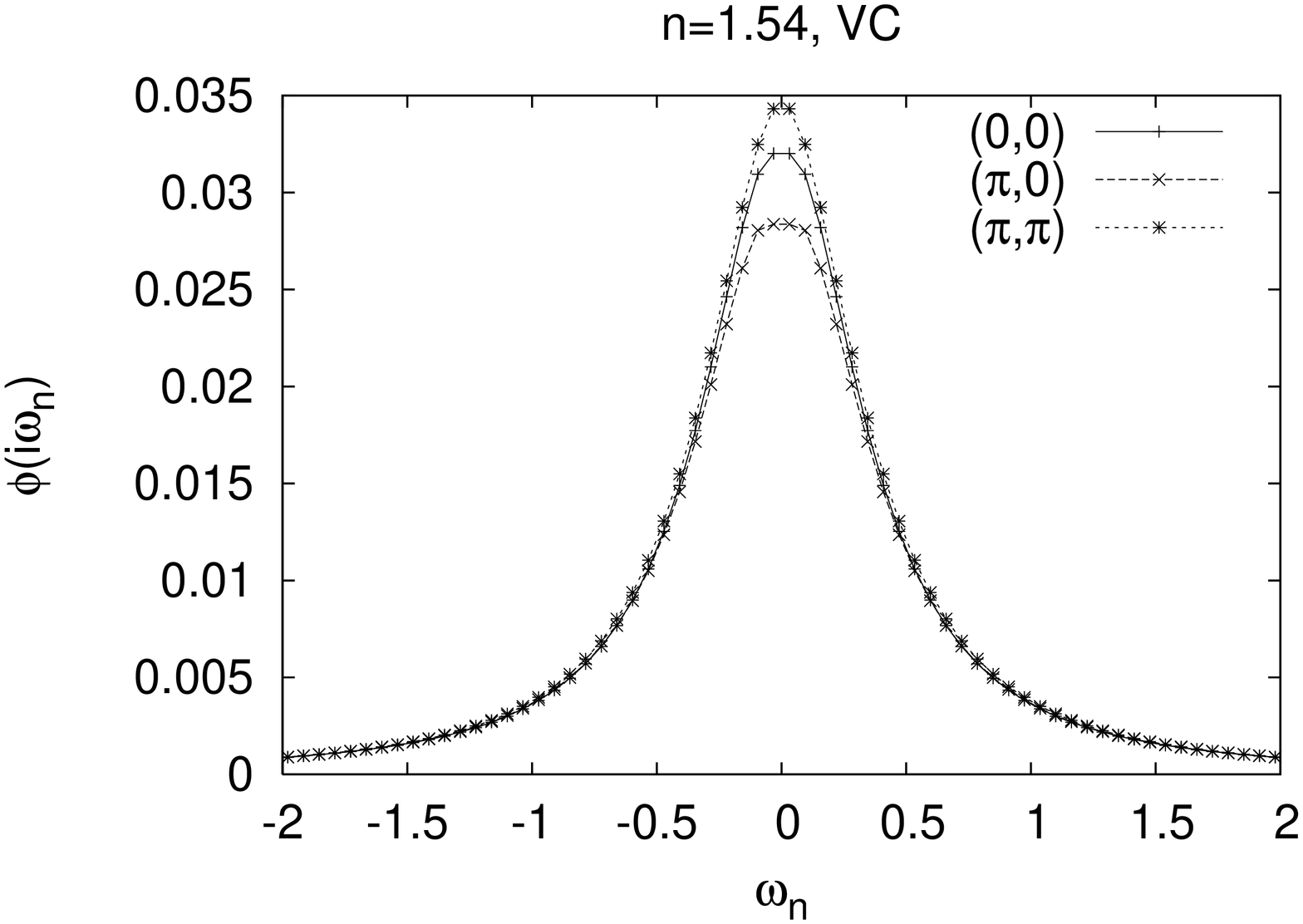}
\end{indented}
\caption{Real part of the anomalous self-energy at various fillings:
  (a) $n=1.12$, no vertex corrections (b) $n=1.54$, no vertex
  corrections (c) $n=1.12$, vertex corrections, (d) $n=1.54$, vertex
  corrections. Calculations were carried out at $T=0.005$ with $U=0.6$
  and $\omega_0=0.4$. Momentum dependence corresponding to non-local
  corrections is clearly visible at half-filling, but drops off as the
  edge of the superconducting phase is reached. Vertex corrections are
  also most important at half-filling. There is a dip in the anomalous
  self-energy because the vertex corrections drop off more quickly in
  $\omega_n$, with opposite sign to $\phi_{\mathrm{ME}}$, indicating
  that the approximation is close to breakdown at half-filling. The
  slight increase in the anomalous self-energy at $n=1.54$ due to
  vertex corrections arises from a change in the form of the electronic
  Green's function. For half filling, the Green's function at the
  van-Hove points is pure imaginary, whereas for the dilute system, it
  is mostly real, so sums over products of Green's functions in the vertex can
  change sign.}
\label{fig:selfenergy}
\end{figure}

It is possible to see the generic effects of vertex and non-local
corrections by examining the anomalous self energy. In figure
\ref{fig:selfenergy}, the anomalous self energy is shown for $n=1.12$
and $n=1.54$ for a cluster size of $N_C=4$ with parameters of $U=0.6$,
$T=0.005$ and $\omega_0=0.4$ with and without vertex corrections. The
panels are (a) $n=1.12$, no vertex corrections (b) $n=1.54$, no vertex
corrections (c) $n=1.12$, vertex corrections, (d) $n=1.54$, vertex
corrections. In panel (a), the momentum dependence of the vertex
neglected theory is clearly visible, and $\phi(\mathbf{K},i\omega_n)$
has a much larger value at the $(\pi,0)$ point. Momentum dependence is
significantly reduced as the system moves away from half-filling
(panel b), indicating that Midgal-Eliashberg theory is more accurate
in dilute systems. This is expected, since in very dilute systems, the
electron density is sufficiently low that electrons meet very
infrequently, and therefore the crossed diagrams of figure
\ref{fig:feynmandiag}b have extremely small contributions. By scanning
vertically, the effect of including vertex corrections can be
seen. Corrections are strongest close to half-filling, and drop off as
the edge of the superconducting phase is reached. Migdal--Eliashberg
theory is clearly quite accurate for dilute 2D systems, but it
consistently fails close to half-filling.  Initially, it seems as
though vertex corrections are larger than the vertex neglected results
at $n=1.12$. In fact, this is not the case. As discussed in
ref. \cite{hague2003a}, at half-filling vertex corrections act to
reduce the magnitude of the phonon self-energy, so there is much less
renormalisation of the phonon propagator. A smaller phonon propagator
means that the effective coupling is smaller, stabilising the
expansion in $\lambda_{\mathrm{eff}}$. There is a dip in the anomalous
self-energy because the vertex corrections drop off more quickly in
$\omega_n$, with opposite sign to $\phi_{\mathrm{ME}}$, which is an
indication that the approximation is close to breakdown at
half-filling. The slight increase in the anomalous self-energy at
$n=1.54$ due to vertex corrections comes about from a change in form
of the electronic Green's function. For half filling, the Green's
function at the van-Hove points is pure imaginary, whereas for the
dilute system, it is mostly real, so sums over products of Green's
functions can change sign with respect to the Migdal--Eliashberg
result. This sign change is also seen in DMFT simulations of the 3D
Holstein model \cite{miller1998a}.

%
At this stage, it is appropriate to examine the size of the parameter
$\lambda_{\mathrm{eff}}$ that defines the vertex correction
expansion. Since the expansion in this case is in the full phonon
Green's function, the expansion parameter is renormalised by the
phonons, and reads
$\lambda_{\mathrm{eff}}=U{\mathcal{D}}(\mu)D(i\omega_s=0)$, where
$D(i\omega_s=0)$ is the phonon propagator at zero Matsubara
frequency. For dilute systems, ${\mathcal{D}}(\mu)$ is typically
small, and so $\lambda_{\mathrm{eff}}$ is small (N.B. Unlike in 3D,
${\mathcal{D}}(\mu)$ is never zero in 2D, because of the discontinuity
in the band edge of the non-interacting DOS). Close to half-filling,
the DOS in 2D is divergent, and this parameter is expected to be
large. In the current approximation, a small interplanar hopping was
applied to stabilise the solution, so $\lambda_{\mathrm{eff}}$ is
smaller than expected in a pure 2D system. As noted in
ref. \cite{hague2003a}, $D(0)$ is reduced by vertex corrections as
compared to the Migdal--Eliashberg result. For most energies, the bare
density of states in 2D is smaller than the bare density of states in
3D, since the divergence drops off logarithmically close to half
filling. Therefore, $\lambda_{\mathrm{eff}}$ is only really large for
$n=1$ within the current parameter range. For the mid to dilute
limits, the relative magnitude of the second order vertex correction
goes like $\lambda^2 \sim 0.04$. At half filling, with the current
parameters, $\lambda^2 \sim 0.5$, so the approximation can only be
considered to be qualitatively correct. Nonetheless, the current
approximation has features appropriate to Hohenberg's theorem
(discussed later) and the bare DOS drops off so quickly moving away
from half-filling, that results are expected to be accurate for most
$n$.

\begin{figure}
\begin{indented}\item[]
\includegraphics[width=35mm,angle=270]{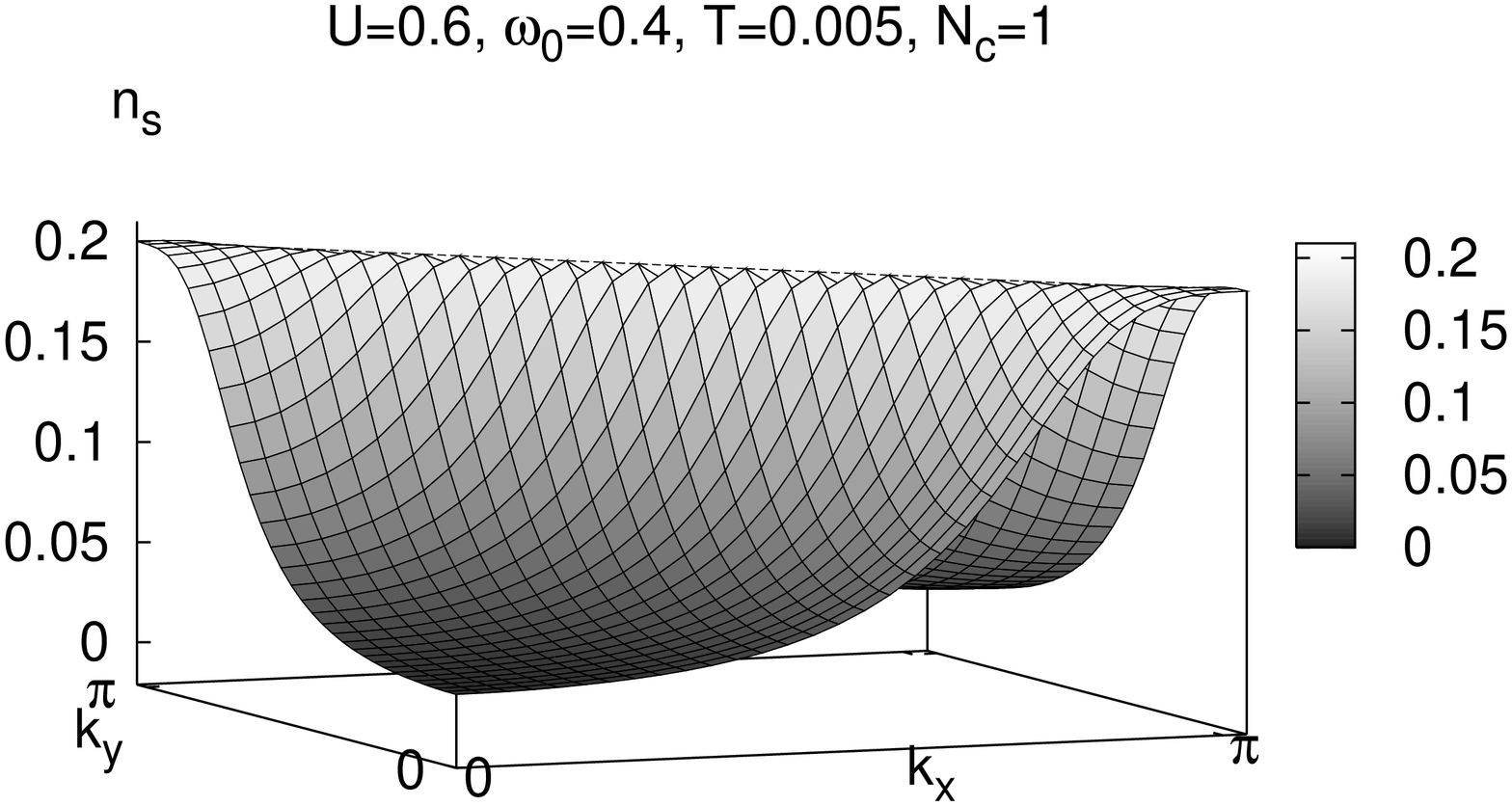}
\includegraphics[width=35mm,angle=270]{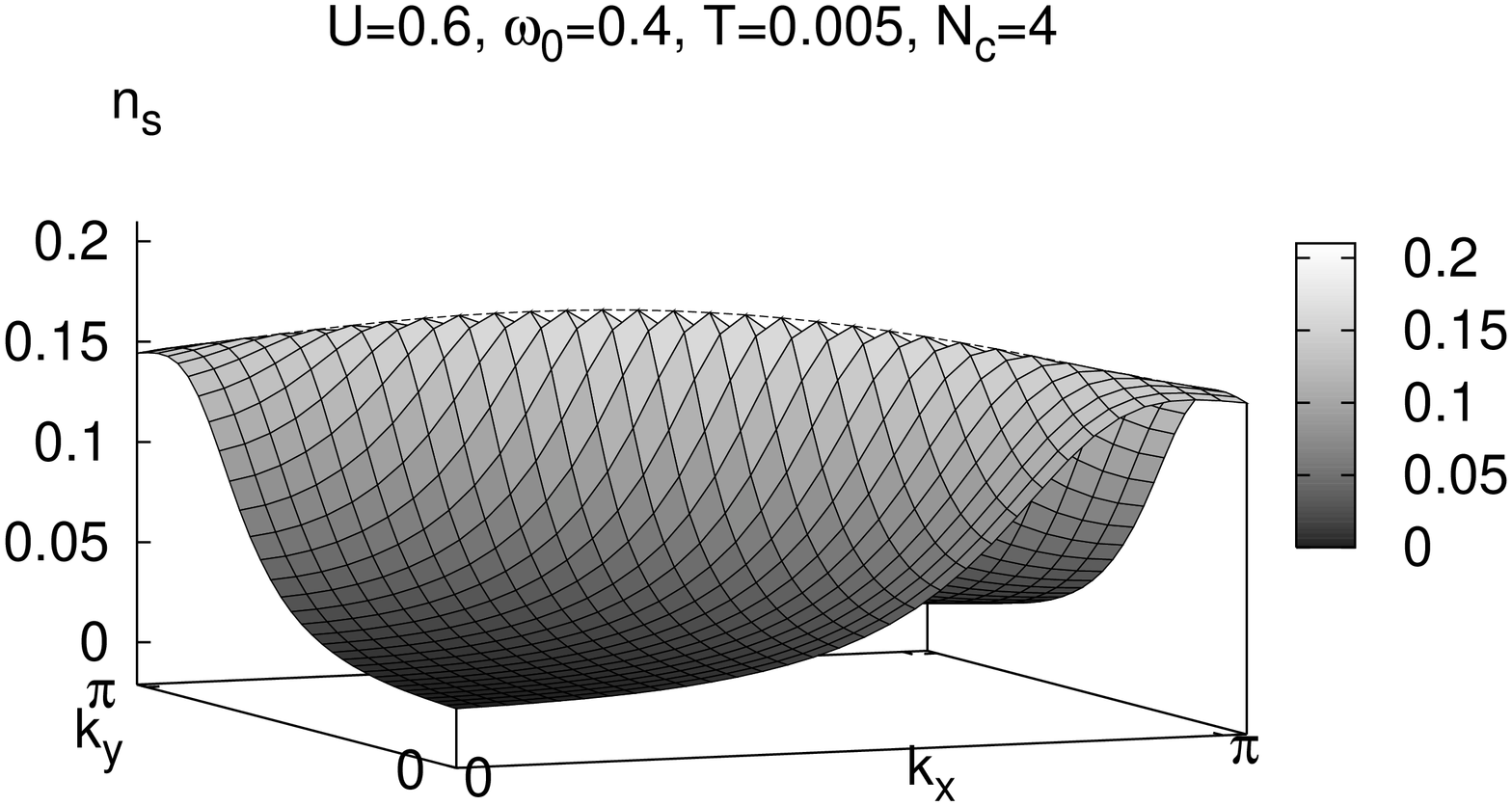}
\includegraphics[width=35mm,angle=270]{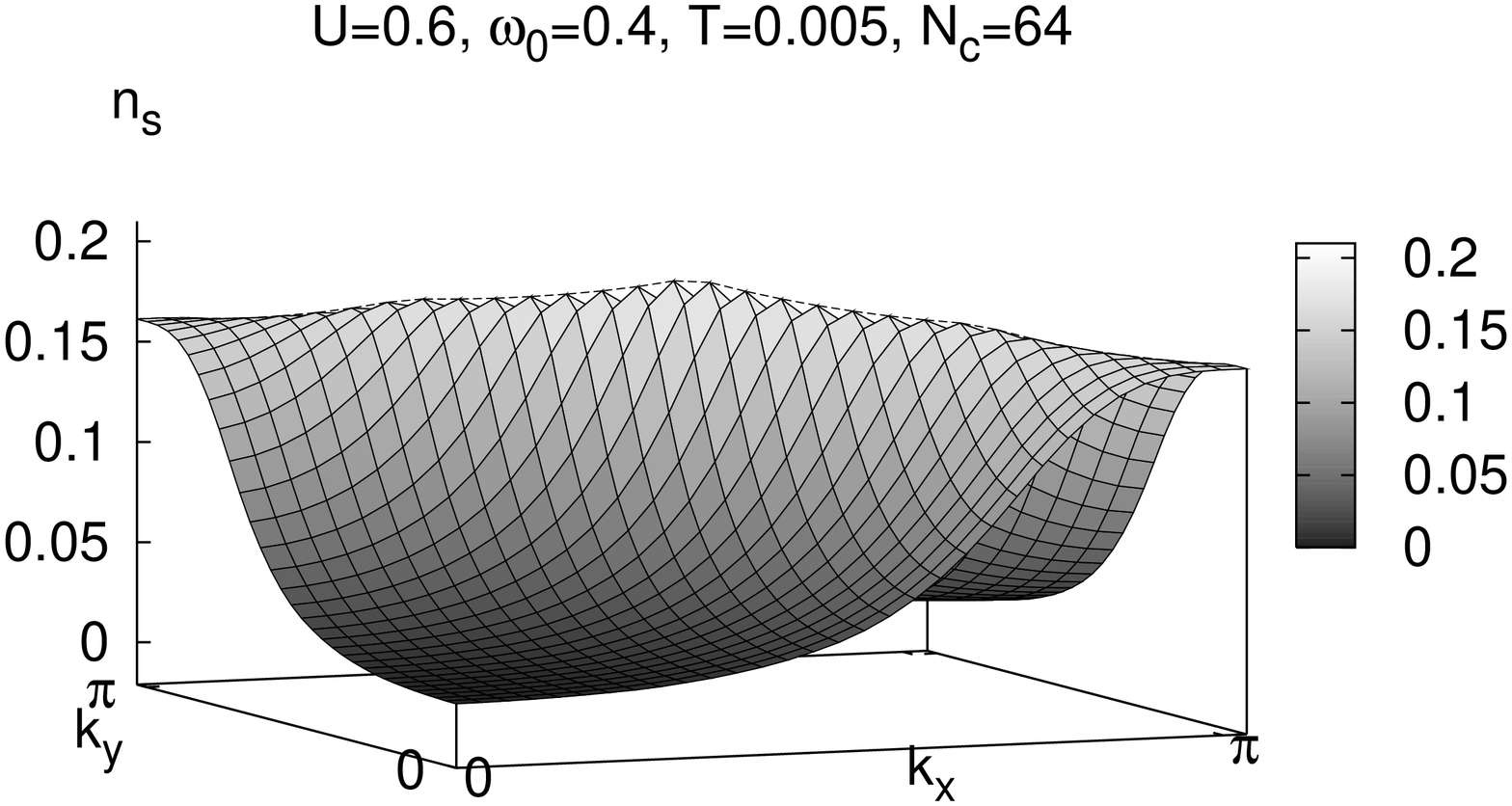}
\includegraphics[width=35mm,angle=270]{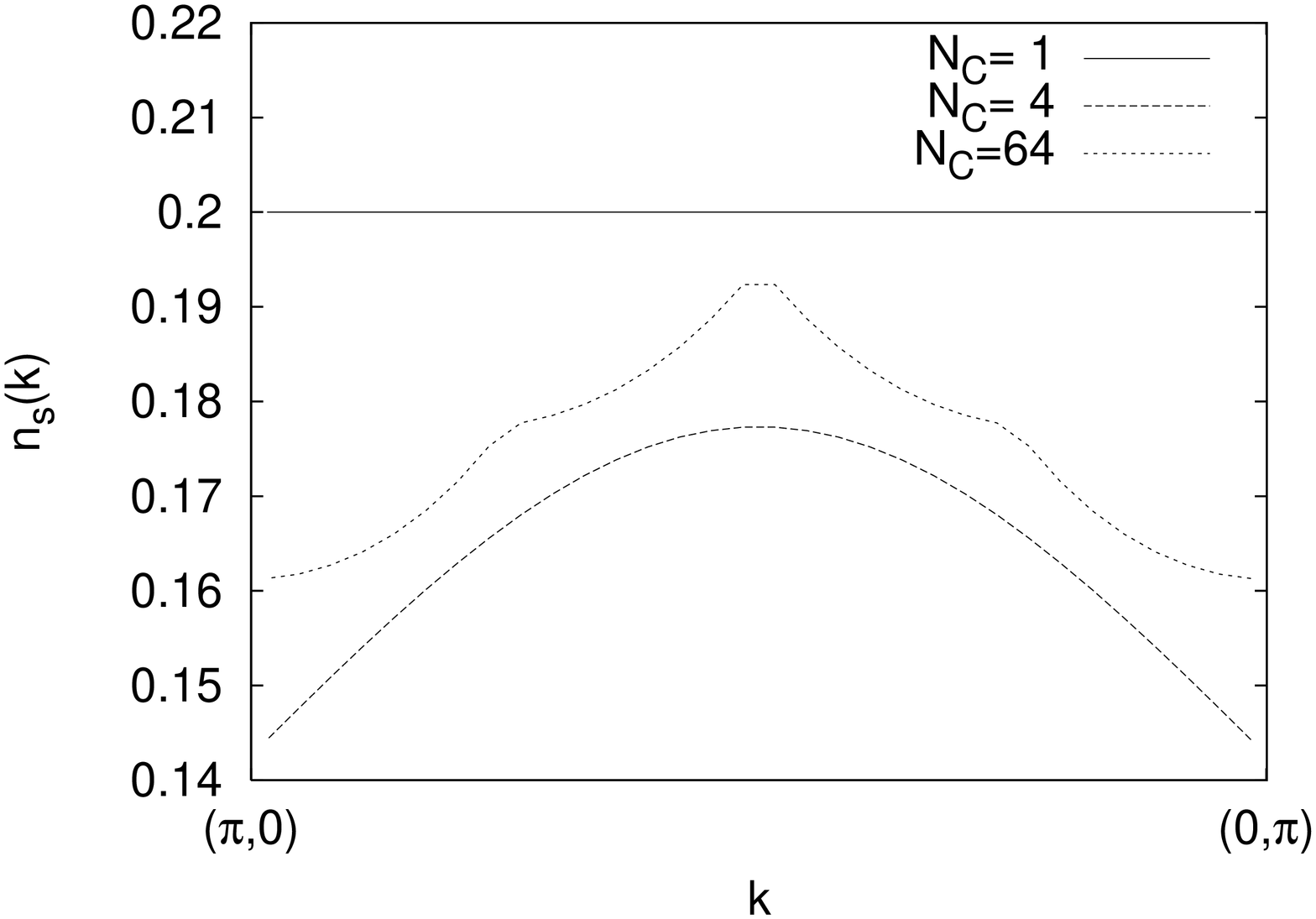}
\end{indented}
\caption{Variation of superconducting (anomalous) pairing density
  across the Brillouin zone. $U=0.6, \omega_0=0.4, n=1$ and
  $T=0.005$. Cluster sizes are increased from $N_C=1$ to
  $N_C=64$. Pairing occurs between electrons close to the
  ``Fermi-surface'' at $\mathbf{k}$ and the opposite face of the
  surface at $-\mathbf{k}$. Also shown is the pairing density at the
  Fermi surface for the 3 different cluster sizes (bottom right). For
  a cluster size of $N_c=1$ corresponding to DMFT, the pairing is
  uniform around the ``Fermi-surface'', demonstrating that momentum
  dependence has been neglected. Momentum dependence favours
  additional pairing along the $k_x=k_y$ line, and a peak can clearly
  be seen. The expansion in spherical harmonics contains even momentum
  states with $m=0$. For $N_c=64$, additional peaks can be seen,
  suggesting that the order parameter also contains extra higher-order
  harmonics. The Fermi-surface is not very clearly defined, with the
  mobile electrons spread out over a significant range of momentum
  states. $T=0.005 << W$, so the spread should be very small in all
  locations in the Brillouin zone, except at the van Hove points,
  $(\pi,0)$ and $(0,\pi)$, indicating that the spreading is due to the
  low dimensionality.}
\label{fig:effectcluster}
\end{figure}

\begin{figure}
\begin{indented}\item[]
\includegraphics[width=70mm,angle=270]{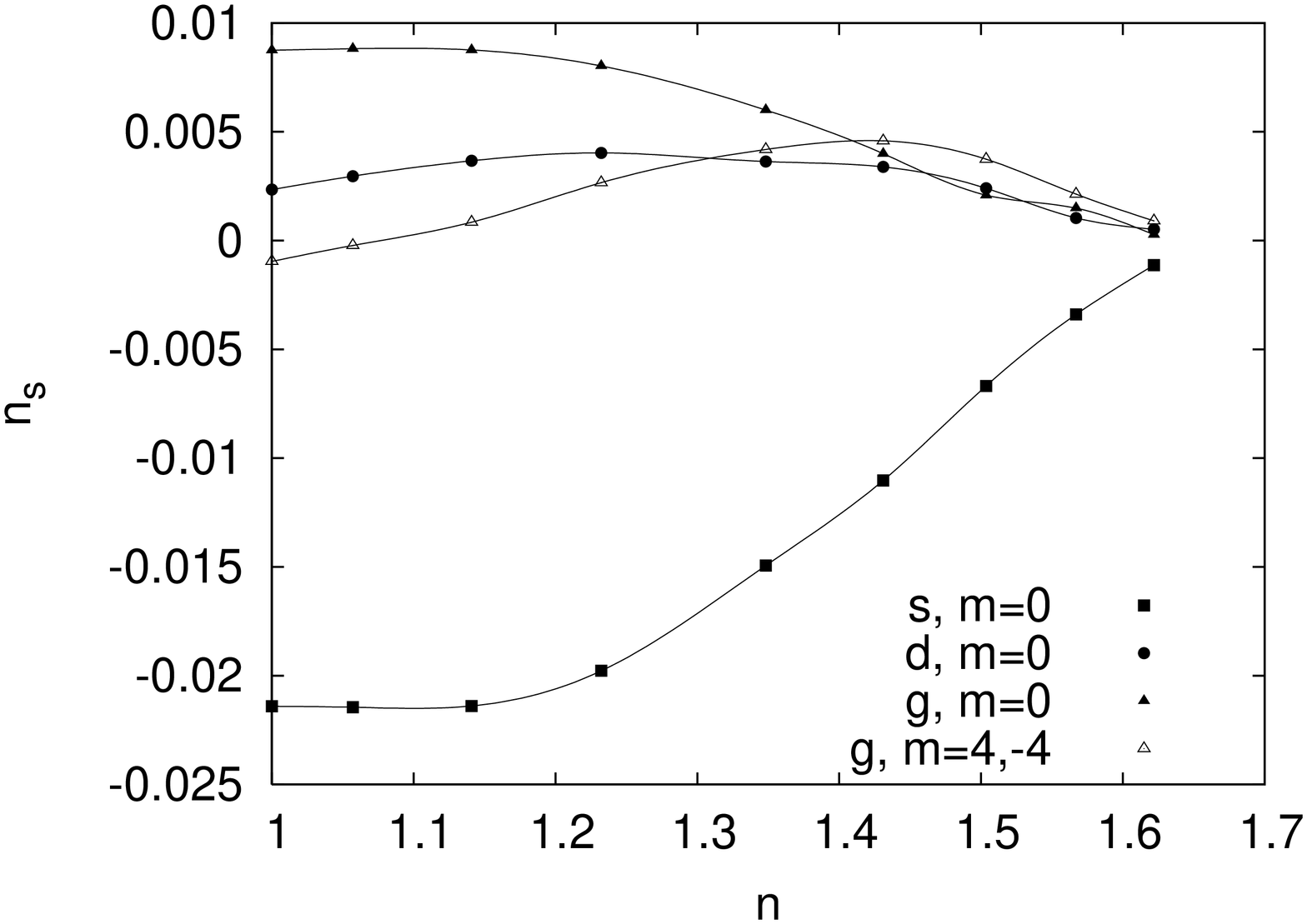}
\end{indented}
\caption{Harmonic decomposition of the anomalous density computed for
  $N_c=64$ as chemical potential is varied. It can be seen that pure
  $s$-wave states are the largest contributors to the anomalous
  density, followed by $g$ and then $d$ states with $m=0$. Owing to
  the hump at the $(\pi/2,\pi/2)$ point, there are also $g$ states
  with $m=\pm 4$. The $m=\pm4$ states have equal magnitude, so
  $m_{\mathrm{tot}}=0$. Note that the relative contribution of higher
  harmonics is greatest away from half filling.}
\label{fig:harmonics}
\end{figure}

How do the differences in the self-energy relate to observable
quantities? One of the big questions in unconventional
superconductivity concerns the possible forms that the order parameter
can take, and a large discussion has grown up around issues such as
the existence of unconventional order parameters such as extended
$s$-wave and higher harmonics. To examine this idea, I demonstrate the
evolution of the shape of the anomalous pairing density
($n_s(\mathbf{k})=|T\sum_n F(\mathbf{k},i\omega_n)|$), which is
related to the order parameter. In this paper, the superconducting
order parameter is treated in a fully self-consistent manner within
the lattice symmetry and no assumptions have been made in advance
about its form.

Figure \ref{fig:effectcluster} shows the variation of superconducting
pairing across the Brillouin zone. In all of the panels, $U=0.6,
\omega_0=0.4, n=1$ and $T=0.005$. A range of cluster sizes is
shown. In the dynamical mean-field theory which corresponds to the
Eliashberg solution (cluster size of $N_c=1$) the pairing is uniform
around the Fermi-surface, as is expected when momentum-dependence is
neglected. The inclusion of non-local momentum dependent fluctuations
has a small, but significant effect on the ordering. Pairing is
reduced most at the $(\pi,0)$ and $(0,\pi)$ points, leading to a
visible peak at $(\pi/2,\pi/2)$. This demonstrates that the order
parameter must necessarily include higher harmonics. For $N_c=64$
additional peaks are also seen. The additional features may be
examined by determining the parameters of an expansion in spherical
harmonics, $c_{lm}=\int \frac{d^{3}\mathbf{k}}{(2\pi)^3}
Y_{lm}(\theta,\phi)n_s(\mathbf{k})$ (figure \ref{fig:harmonics}). This
shows that the anomalous density can be thought of as $m=0$ harmonics
with $s,d,g ...$ character, and additional harmonics with $m=\pm 4$ in
the $g$ channel. The harmonics can be quite large, especially away
from half-filling, and undoubtedly need to be included if the
superconductivity is to be described correctly.

\begin{figure}
\begin{indented}\item[]
\includegraphics[width=35mm,angle=270]{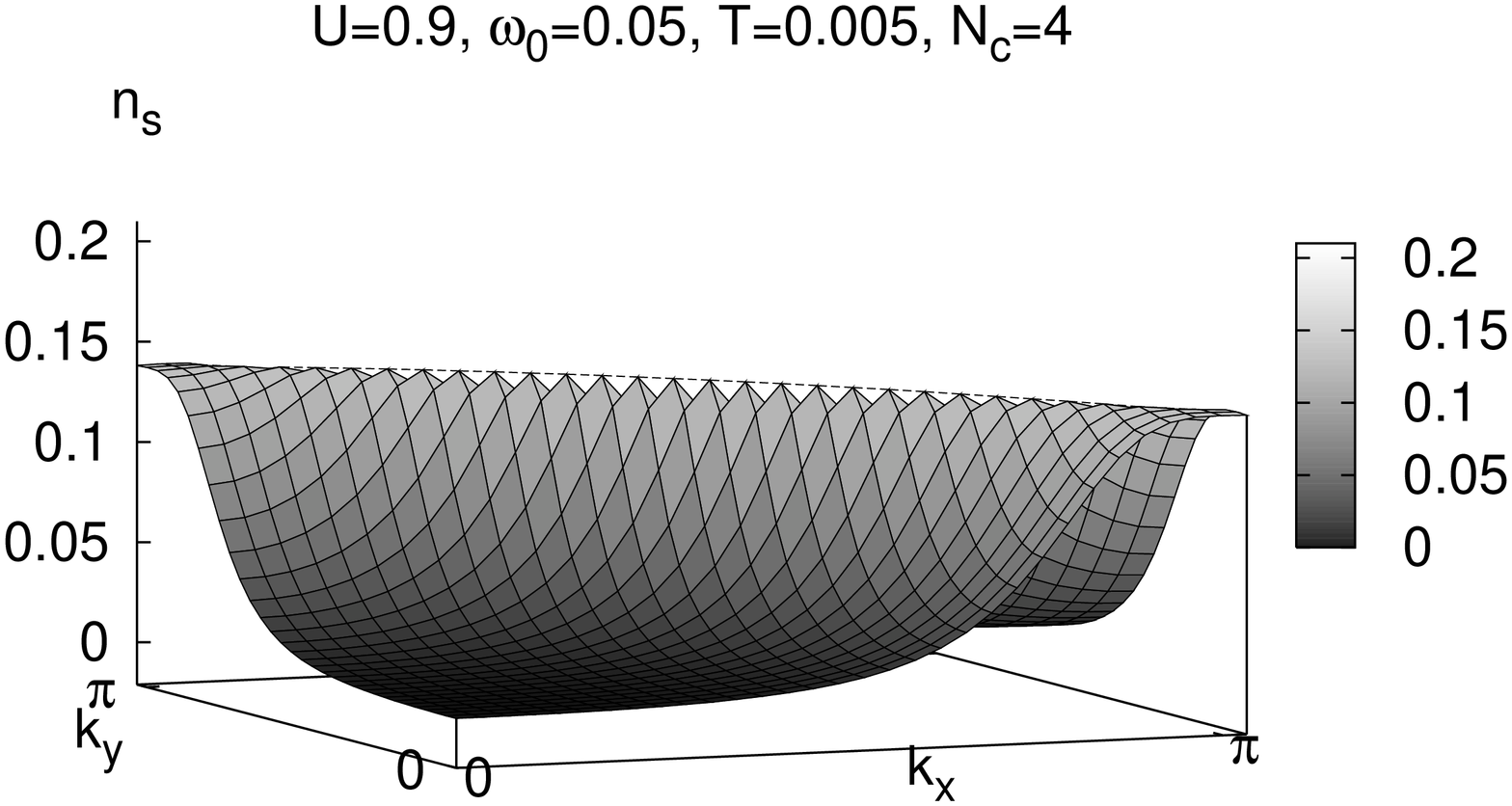}
\includegraphics[width=35mm,angle=270]{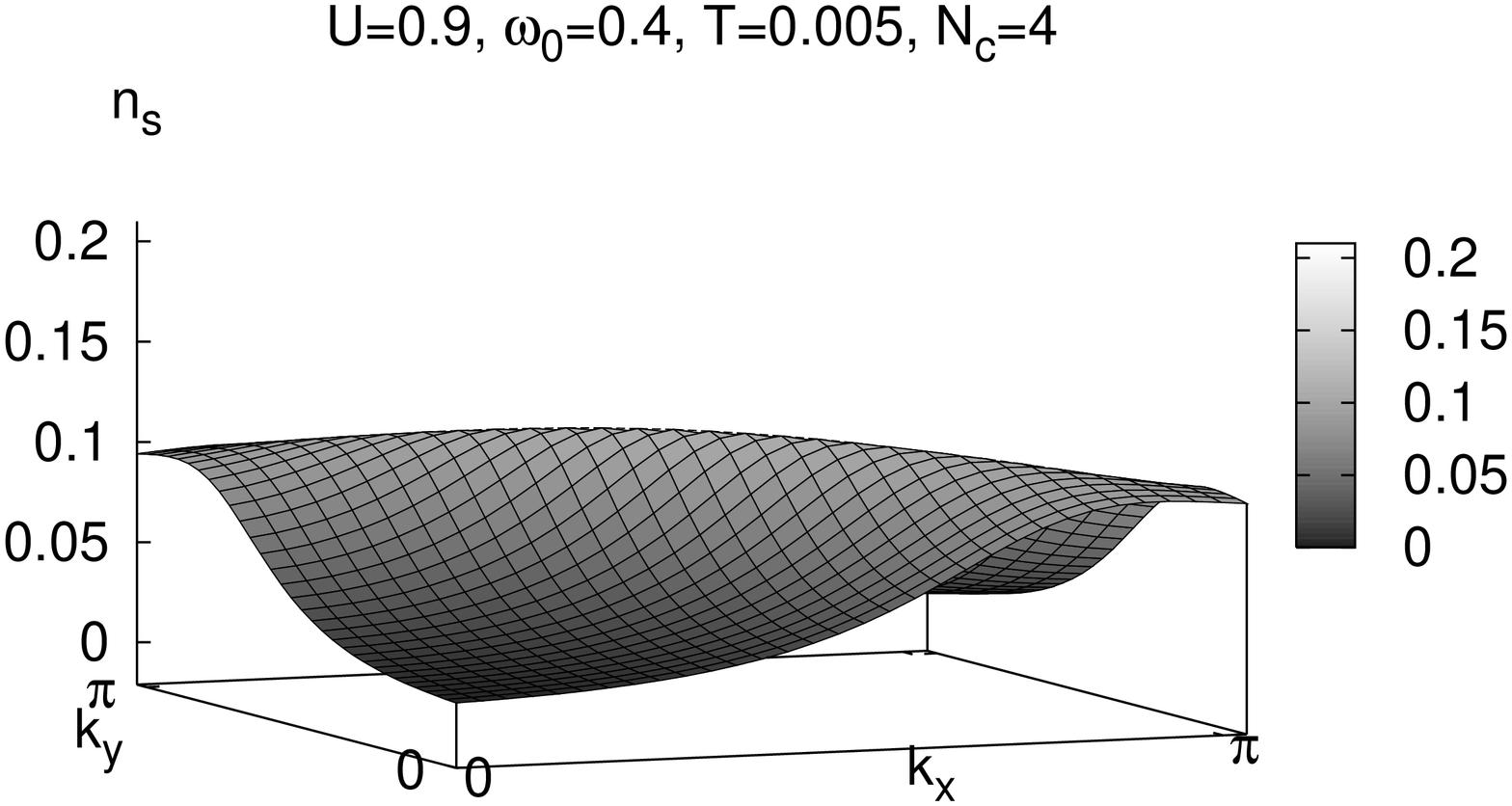}
\includegraphics[width=35mm,angle=270]{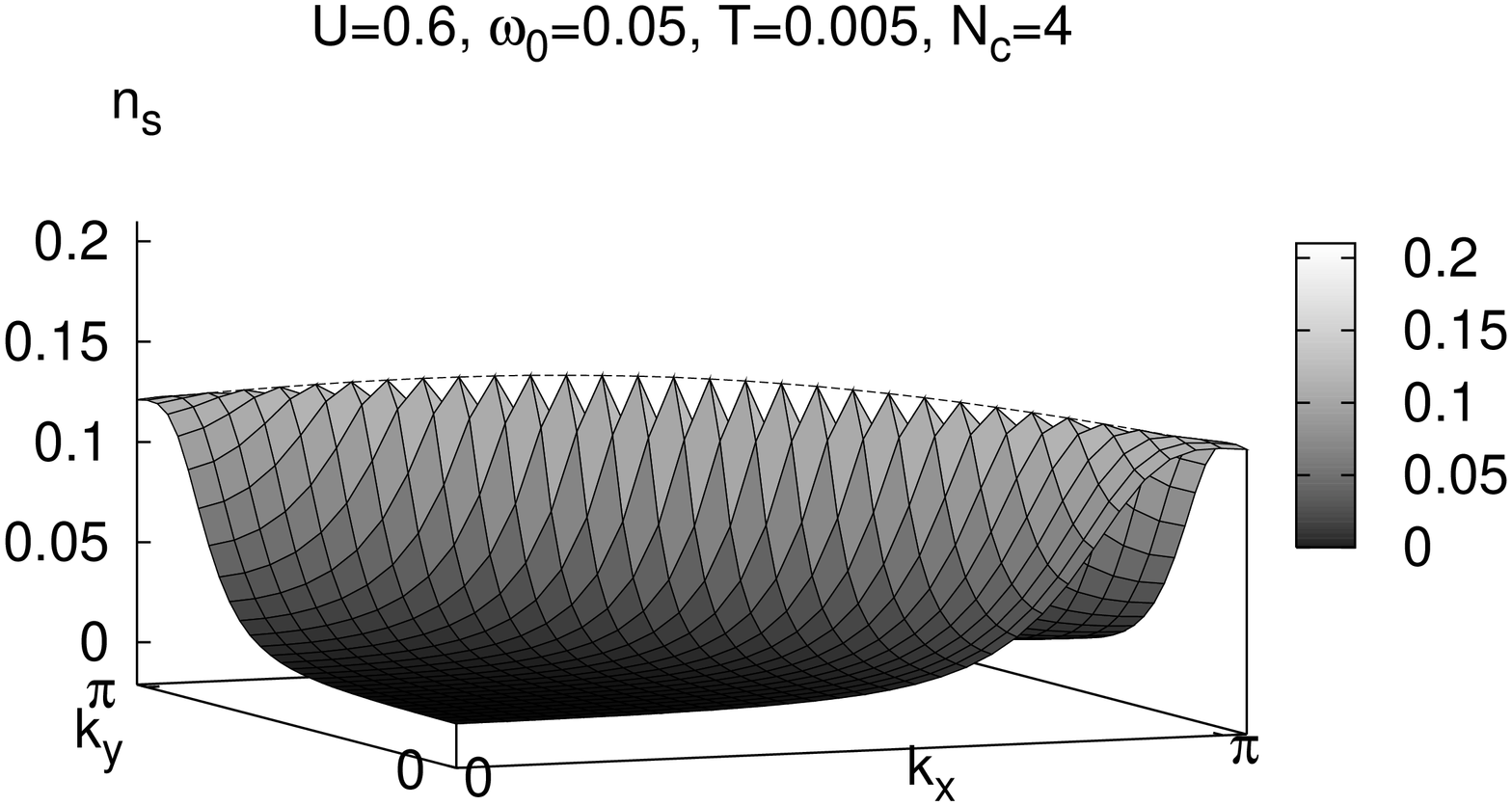}
\includegraphics[width=35mm,angle=270]{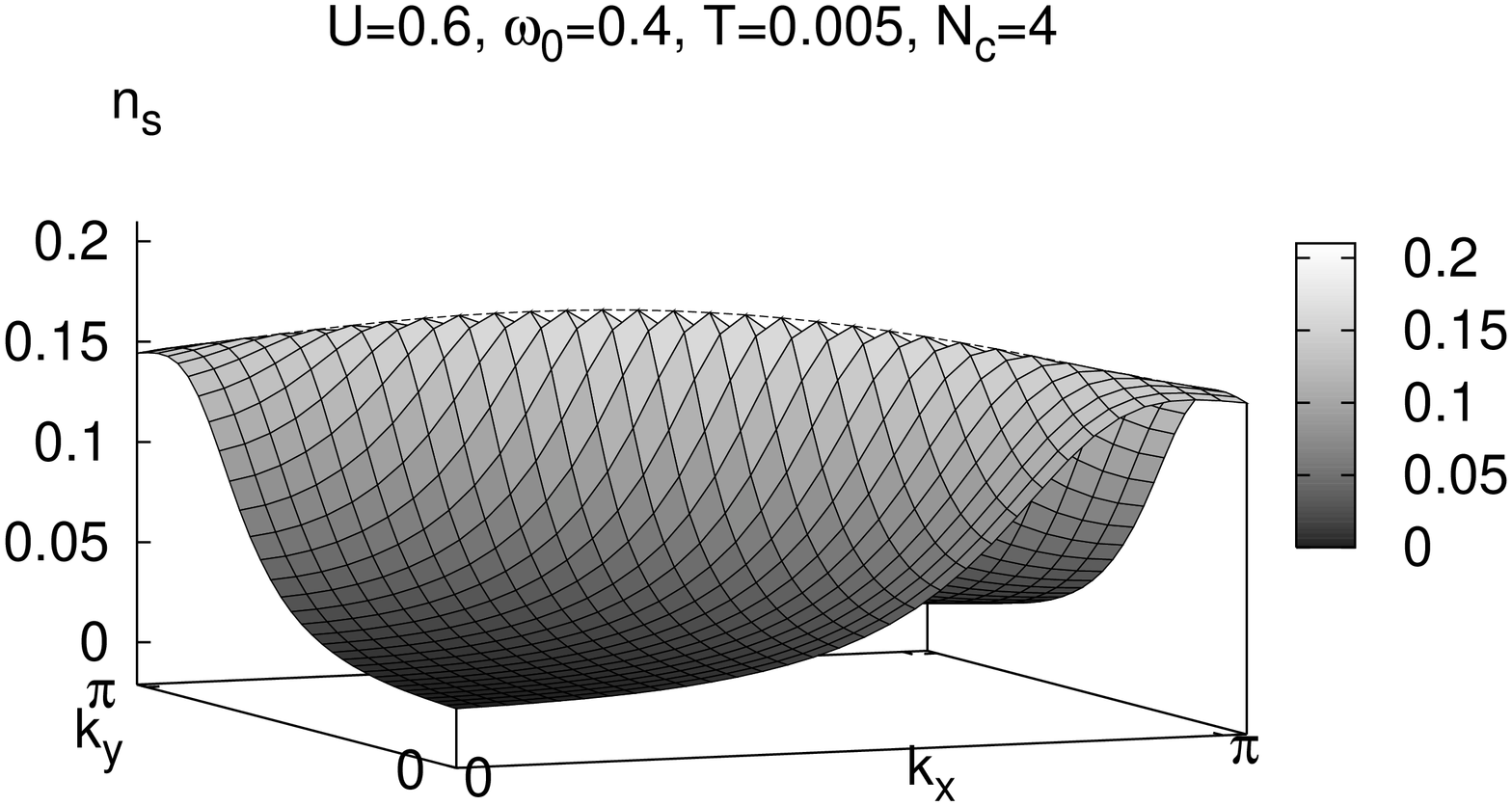}
\end{indented}
\caption{Variation of superconducting (anomalous) pairing density
across the Brillouin zone. $T=0.005$ and $N_C=4$. Changes in the order
parameter are shown as coupling and phonon frequency are changed. As
the phonon frequency is increased, the momentum dependence also
increases, and the Fermi-surface is less well defined. For $U=0.9$,
$\omega_0=0.05$, the order parameter is almost flat along the
Fermi-surface, indicating that DMFT is a good approximation for those
parameters. For the largest coupling and phonon frequencies (at the
edge of applicability for the current approximation), the Fermi-surface
is practically destroyed.}
\label{fig:effectcouple}
\end{figure}


Figure \ref{fig:effectcouple} shows the variation of superconducting
(anomalous) pairing density across the Brillouin zone as coupling and
phonon frequency are changed. $T=0.005$, $n=1$ and $N_C=4$ with vertex
corrections excluded. As the phonon frequency is increased, the
momentum dependence also increases. For $U=0.9$, $\omega_0=0.05$, the
order parameter is almost flat along the Fermi-surface, indicating
that DMFT is a good approximation for those parameters. Typically,
additional coupling makes the order more uniform in the Brillouin
zone. Note that for very strong coupling, the DMFT solution is
expected to become exact, even in 2D, since the bare dispersion is
then essentially flat, and the problem is completely local. For weak
coupling and phonon frequency, there is a well defined
Fermi-surface. For the largest coupling and phonon frequencies (at the
edge of applicability for the current approximation), the
Fermi-surface is practically destroyed.

\begin{figure}
\begin{indented}\item[]
\includegraphics[width=70mm,angle=270]{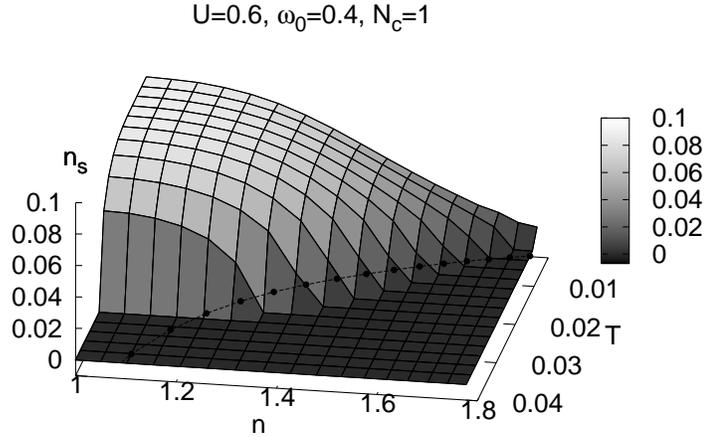}
\end{indented}
\caption{Superconducting phase diagram showing the total number of
  superconducting states. $U=0.6, \omega_0=0.4$ and various $T$. A
  cluster size of $N_c=1$ has been used, and no vertex corrections are
  included. The superconductivity is strongest at half-filling and the
  order drops off monotonically as the filling increases. Results in
  the dilute limit are in good agreement with the BCS result (the
  transition temperature from BCS is shown as the line with points in
  the $n_s=0$ plane). Closer to half-filling, the DMFT result is
  significantly smaller than the BCS result (which predicts
  $T_C(n=1)>0.07$). The difference in results between the two
  mean-field theories at half-filling is due to the self-consistency
  in the DMFT.}
\label{fig:phasediagram1}
\end{figure}

%
It is of clear interest to map the phase diagram associated with
superconducting order. First, the superconductivity arising from DMFT
is investigated in the absence of vertex corrections. Figure
\ref{fig:phasediagram1} shows the resulting phase diagram. Note that
in the DMFT solution, the superconductivity is strongest at
half-filling and the order drops off monotonically as the filling
increases. Assuming a form for the density of states in 2D (with small
interplane hoppnig) of
${\mathcal{D}}(\epsilon)=(1-t\log[(\epsilon^2+t_{\perp}^2)/16t^2])/t\pi^2$
(for $|\epsilon|<4t$) \cite{liliana}, the BCS result may be calculated
using the expression
$T_C(n)=2\omega_0\exp(-1/|U|{\mathcal{D}}(\mu(n)))/\pi$, with the
chemical potential taken from the self-consistent solution for a given
$n$. This result also drops off monotonically. Results in the dilute
limit are in good agreement with the BCS result. Closer to
half-filling, the DMFT result is significantly smaller than the BCS
result (which predicts $T_C(n=1)>0.07$). The difference in results
between the two mean-field theories at half-filling is due to the
self-consistency in the DMFT. For small $U$, the self-consistent
equations converge on the first iteraction, but for larger $U$, the
phonon and electron Green's functions are significantly renormalised,
thus reducing the transition temperature.

\begin{figure}
\begin{indented}\item[]
\includegraphics[width=70mm,angle=270]{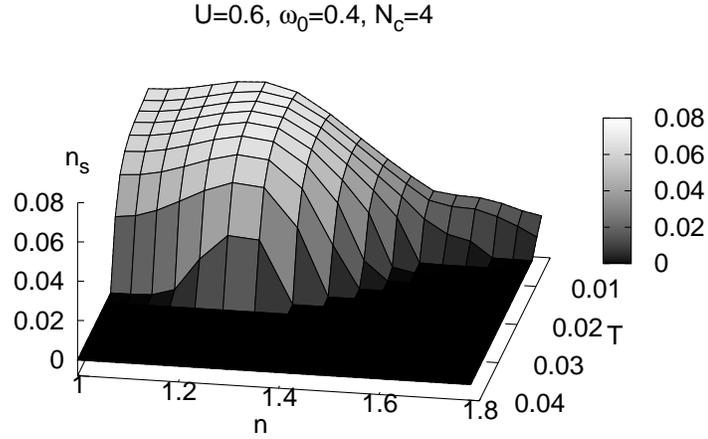}
\end{indented}
\caption{Superconducting phase diagram showing the total number of
  superconducting states. $U=0.6, \omega_0=0.4$ and various $T$. A
  cluster size of $N_c=4$ has been used, and no vertex corrections are
  included. There is an anomalous bump centred about $n=1.25$,
  indicating that the strongest superconductivity occurs away from
  half filling. The highest transition temperature occurs for
  $T=0.025$. The reduction in the transition temperature close to
  half-filling shows the onset of Hohenberg's theorem. The largest
  superconductivity coincides with the increase in components without
  pure $s$-wave character (see figure \ref{fig:harmonics}).}
\label{fig:phasediagram4}
\end{figure}

%
To show the differences induced by spatial fluctuations, the phase
diagram is computed for a cluster size of $N_C=4$. Figure
\ref{fig:phasediagram4} shows the total density of superconducting
pairs for $U=0.6, \omega_0=0.4$ and various temperatures and fillings,
without vertex corrections. Of most interest is an anomalous bump
centred about $n=1.25$, indicating that the strongest
superconductivity occurs away from half filling, and that this is due
to non-local fluctuations in 2D. Assuming a material with a
non-interacting band width of 1eV, the highest transition temperature
would correspond to 145K. This value is higher that that eventually
expected in real materials. For instance, the effect of a Coulomb
pseudopotential $U_C$ will be a reduction of the transition
temperature. The standard BCS result is modified by Coulomb repulsion
in the following way, $T_C = 2\omega_0 \exp(-1/(\lambda-U_C)))/\pi$.

%
In addition to the $T_C$ reduction due to Coulomb repulsion, a
fundamental limit on the transition temperature in pure 2D is the
Hohenberg theorem \cite{hohenberg}. This is closely related to the
effects of spatial fluctuations. The basis of Hohenberg's proof is the
divergence of certain quantities (which are known to be finite) in
$d\le 2$ at $\mathbf{k}=0$ when anomalous expectation values (e.g. the
superconducting order parameter) are non zero. In the DMFT solution,
there are no specific $\mathbf{k}=0$ states due to the coarse
graining, and so the divergence in the correlation functions that led
Hohenberg to determine that the order parameter must be zero for zero
momentum pairing in $d\le 2$ is washed out, leading to a
finite transition temperature for the 2D local approximation. In DCA,
partial momentum dependence is restored. Therefore, the effects of the
washed out divergences are stronger. There is still a finite
transition temperature, but it is reduced wherever there is strong
momentum dependence. This is demonstrated by the drop in
superconducting order at and close to half-filling in figure
\ref{fig:phasediagram4}, where the momentum dependence is
strongest. As the number of cluster points increases, the momentum
resolution becomes superior, and the divergences of Hohenberg's
theorem are expected to emerge in a systematic manner. In real
materials with quasi-2D character, some interplane hopping remains. In
that case, the results from small cluster DCA are expected to be more
reliable.

\begin{figure}
\begin{indented}\item[]
\includegraphics[width=70mm,angle=270]{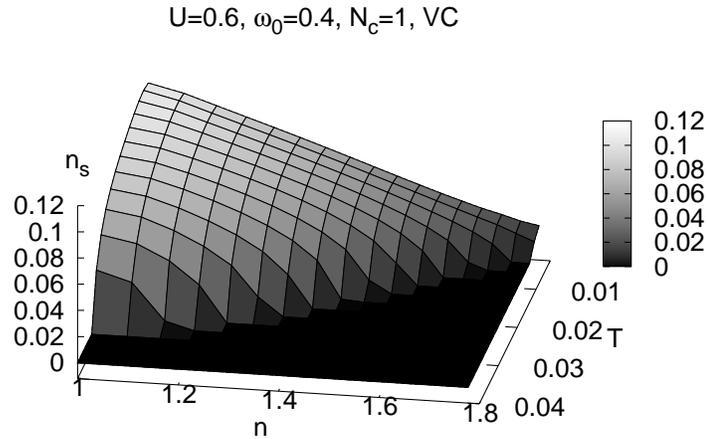}
\end{indented}
\caption{Superconducting phase diagram showing the total number of
superconducting states. $U=0.6, \omega_0=0.4$ and various $T$. A
cluster size of $N_c=1$ has been used, and vertex corrections are
included. As in figure \ref{fig:phasediagram1}, the DMFT result falls
off monotonically with increased filling.}
\label{fig:phasediagram1vc}
\end{figure}

\begin{figure}
\begin{indented}\item[]
\includegraphics[width=70mm,angle=270]{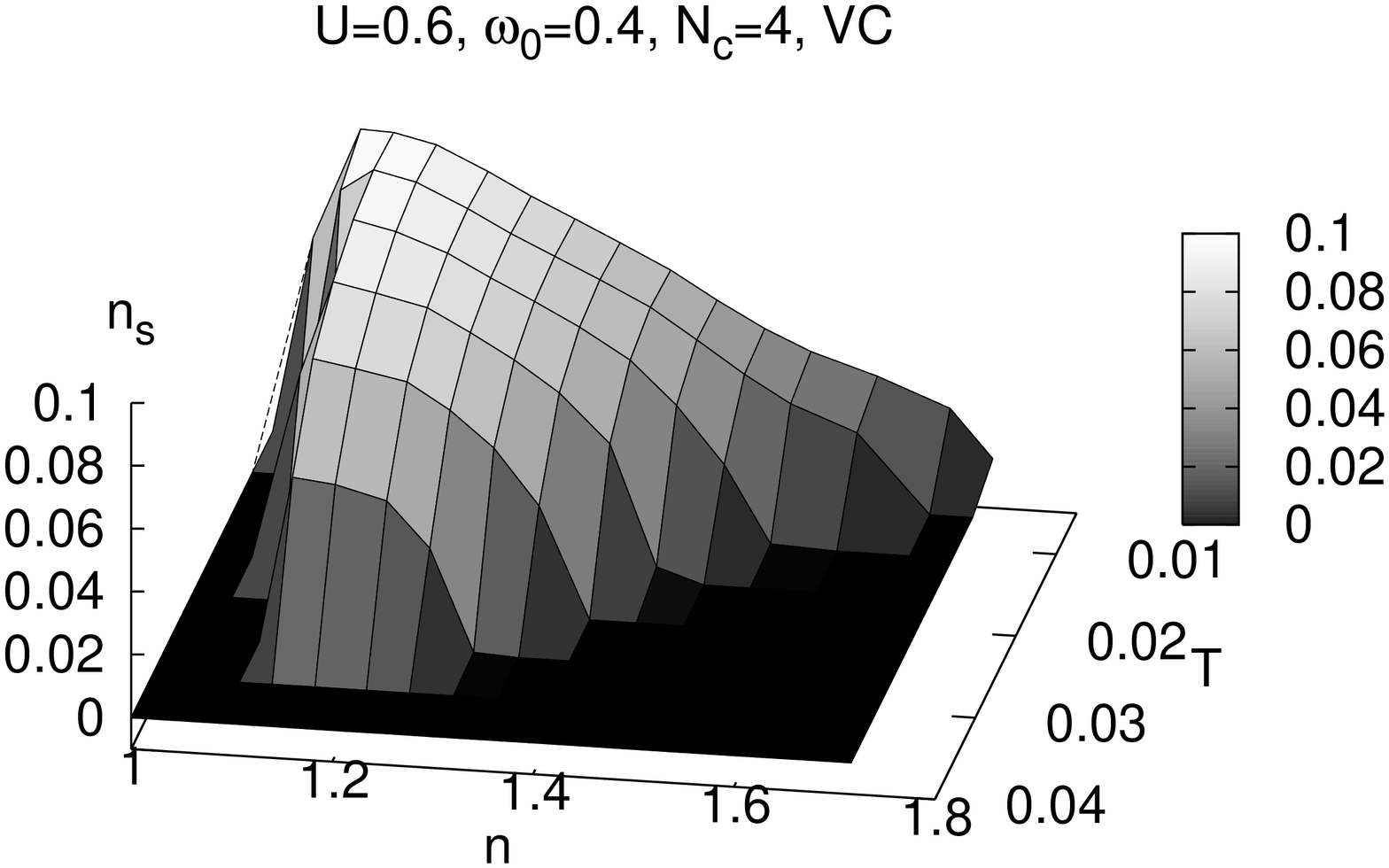}
\end{indented}
\caption{Superconducting phase diagram showing the total number of
  superconducting states. $U=0.6, \omega_0=0.4$ and various $T$. A
  cluster size of $N_c=4$ has been used, and vertex corrections are
  included. Superconducting states are suppressed at half-filling, and
  there is a significant bulge away from half-filling, with a maximum
  transition temperature of 0.015W. It is significant that the
  transition temperature is reduced to zero at half-filling and
  supressed close to half filling, since reduction of transition
  temperatures is expected in 2D due to Hohenberg's theorem. }
\label{fig:phasediagram4vc}
\end{figure}

Finally, I demonstrate the effects of vertex corrections on the
superconducting phase diagram. Figures \ref{fig:phasediagram1vc} and
\ref{fig:phasediagram4vc} show the total number of superconducting
states as a function of filling and temperature for cluster sizes of
$N_C=1$ and $N_C=4$ respectively. An electron-phonon coupling of
$U=0.6$ and phonon frequency of $\omega_0=0.4$ have been used. Vertex
corrections do not appear to make a large difference to the DMFT
result in figure \ref{fig:phasediagram1vc}. For $N_C=4$, the bulge
that was seen in the non vertex-corrected phase diagram is very
clearly enhanced. Superconductivity at half-filling is completely
suppressed in the vertex corrected solution. This is the precursor to
Hohenberg's theorem applying across the entire phase diagram, and both
the effects of spatial fluctuations and the lowest order vertex
correction were essential to obtain that agreement.

\section{Summary}
\label{section:summary}

In this paper I have carried out DCA calculations of a quasi-2D Holstein
model in the superconducting state with large in plane hopping and
small out of plane hopping ($t=0.25, t_{\perp}=0.01$). Several
approximations to the self-energy were considered, including the
neglect of vertex corrections (which corresponds to a
momentum-dependent extension to the Eliashberg theory), the inclusion
of vertex corrections as a corrected approximation for stronger
couplings, and the introduction of spatial fluctuations. The anomalous
self energy, superconducting order parameter and phase diagram were
calculated.

The superconducting order parameter was found to modulate around the
fermi surface, and is not pure $s$-wave. Analysis of the harmonics
showed that the state is a conbination of $s, d, f$ etc. states with
$m=0$, and other states with integer value of $m=\pm4n$. The total
angular momentum is always zero. The contribution of the $m\neq0$
states is considerably larger away from half filling. Increases in
bare phonon frequency tended to increase the strength of the
superconducting order, and contributed to a degeneration of the
Fermi-surface.

The phase diagram was analysed for small cluster sizes of
$N_C=1,4$. For $N_C=1$, the phase diagram was shown to agree
qualitatively with the BCS theory. When spatial fluctuations are
included, the superconducting order is suppressed at half-filling,
leading to a characteristic hump at a doping of approximately $\delta
n=0.25$. Vertex corrections completely suppress superconductivity at
half-filling, which is believed to be a manifestation of Hohenberg's
theorem. In particular, the states with the largest momentum
dependence showed the strongest reduction in the transition
temperature, indicating that spatial fluctuations as well as vertex
corrections contribute to the supression of superconducting order in
pure 2D materials.

\ack

JPH would like to thank the University of Leicester for hospitality
and use of facilities while carrying out this research.

\end{document}